\documentclass[twocolumn,prb,showpacs,preprintnumbers,amsmath,amssymb]{revtex4}
\usepackage{graphicx}
\usepackage{dcolumn}
\usepackage{bm}

\begin{document}

\title{Tunneling-percolation origin of nonuniversality: theory and experiments}
\author{Sonia Vionnet-Menot$^1$}
\author{Claudio Grimaldi$^1$}\email{claudio.grimaldi@epfl.ch}
\author{Thomas Maeder$^{1,2}$}
\author{Sigfrid Str\"assler$^{1,2}$}
\author{Peter Ryser$^1$}
\affiliation{$^1$ Laboratoire de Production Microtechnique, Ecole
Polytechnique F\'ed\'erale de Lausanne, Station 17, CH-1015
Lausanne, Switzerland} \affiliation{$^2$Sensile Technologies SA,
PSE, CH-1015 Lausanne, Switzerland}

\begin{abstract}
A vast class of disordered conducting-insulating compounds close
to the percolation threshold is characterized by nonuniversal
values of transport critical exponent $t$, in disagreement with
the standard theory of percolation which predicts $t\simeq 2.0$
for all three dimensional systems. Various models have been
proposed in order to explain the origin of such universality
breakdown. Among them, the tunneling-percolation model calls into
play tunneling processes between conducting particles which,
under some general circumstances, could lead to transport
exponents dependent of the mean tunneling distance $a$. The
validity of such theory could be tested by changing the parameter
$a$ by means of an applied mechanical strain. We have applied this
idea to universal and nonuniversal RuO$_2$-glass composites. We
show that when $t>2$ the measured piezoresistive response
$\Gamma$, {\it i. e.} the relative change of resistivity under
applied strain, diverges logarithmically at the percolation
threshold, while for $t\simeq 2$, $\Gamma$ does not show an
appreciable dependence upon the RuO$_2$ volume fraction. These
results are consistent with a mean tunneling dependence of the
nonuniversal transport exponent as predicted by the
tunneling-percolation model. The experimental results are compared
with analytical and numerical calculations on a random-resistor
network model of tunneling-percolation.
\end{abstract}

\maketitle

\section{introduction}
\label{intro} Despite that transport properties of disordered
insulator-conductor composites have been studied for more than
thirty years, some phenomena still remain incompletely understood.
One such phenomenon is certainly the origin of nonuniversality of
the DC transport near the conductor-insulator critical transition.
According to the standard theory of transport in isotropic
percolating materials, the bulk conductivity $\sigma$ of a
composite with volume concentration $x$ of the conducting phase
behaves as a power-law of the form:\cite{stauffer,sahimi}
\begin{equation}\label{rho}
\sigma\simeq \sigma_0 (x-x_c)^t,
\end{equation}
where $\sigma_0$ is a proportionality constant, $x_c$ is the
critical concentration below which the composite has zero
conductivity (or more precisely the conductivity of the
insulating phase) and $t$ is the DC transport critical exponent.
The above expression holds true in the critical region $x-x_c \ll
1$ in which critical fluctuations extend over distances much
larger than the characteristic size of the constituents. As a
consequence, contrary to $\sigma_0$ and $x_c$ which depend on
microscopic details such as the microstructure and the mean
inter-grain junction conductance, the exponent $t$ is expected to
be material independent.\cite{stauffer,sahimi} The universality
of $t$ is indeed confirmed by various numerical calculations of
random-resistor network models which have established that
$t=t_0\simeq 2.0$ for three dimensional lattices to a rather high
accuracy.\cite{batrouni,normand,clerc}
\begin{figure}[t]
\protect
\includegraphics[scale=0.58]{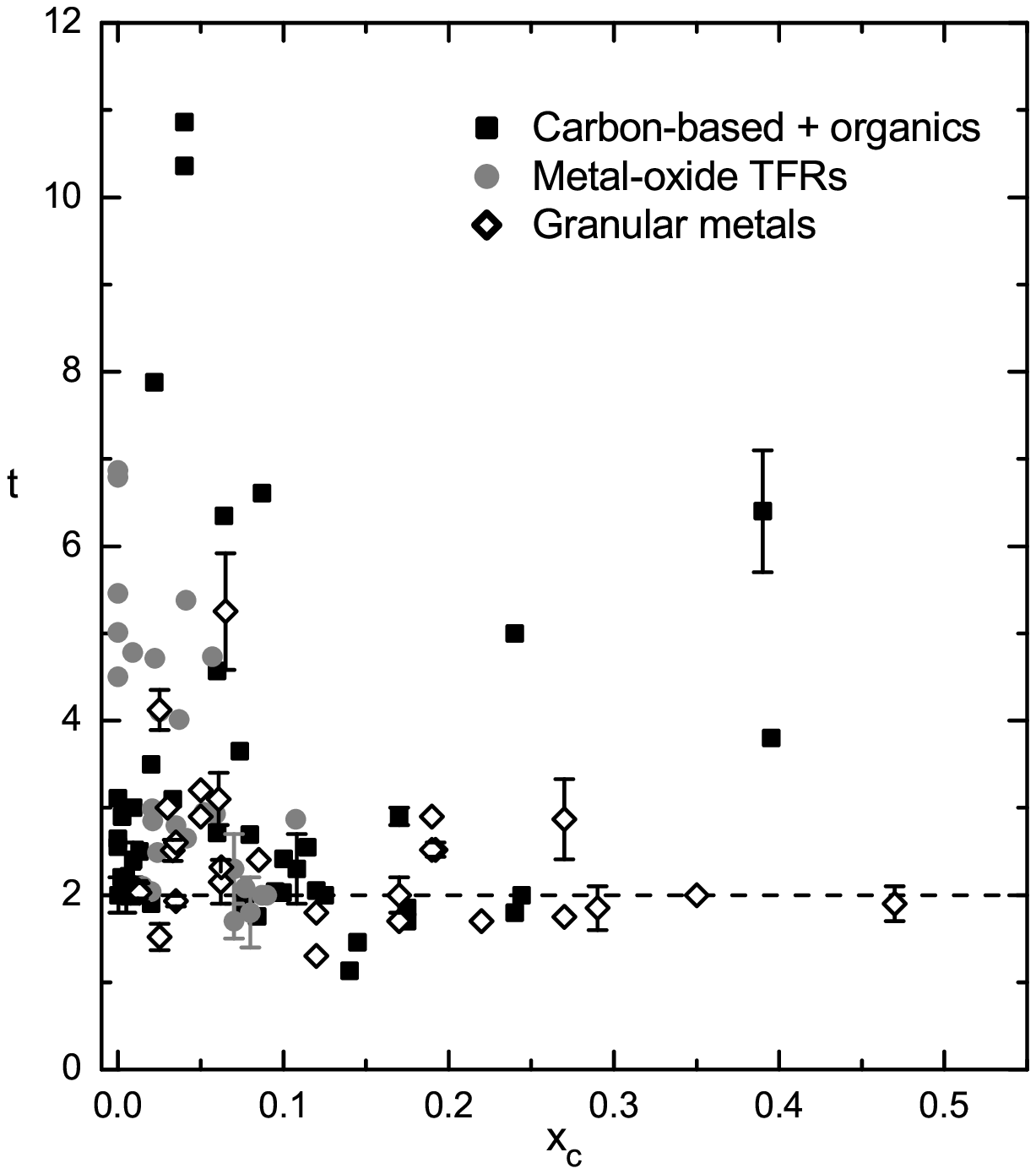}
\caption{Collection of critical exponent values $t$ and
corresponding critical threshold concentration $x_c$ for various
disordered conductor-insulator composites. Carbon-black--polymer
systems are from
Refs.[\onlinecite{flandin1}-\onlinecite{fournier}], oxide-based
thick film resistors are from
Refs.[\onlinecite{pike}-\onlinecite{dziedzic1}], metal-inorganic
and -organic insulator granular metals are from
Refs.[\onlinecite{fournier,chiteme}-\onlinecite{sun}]. The dashed
line denotes the universal value $t_0\simeq 2.0$. } \label{fig1}
\end{figure}

Confirmations to the standard percolation theory of transport
universality are found only in a limited number of experiments on
real disordered composites. This is illustrated in Fig.\ref{fig1}
where we have collected 99 different values of the critical
exponent $t$ and the critical threshold $x_c$ measured in various
composites including carbon-black--polymer
systems,\cite{flandin1}$^-$\cite{fournier} oxide-based thick film
resistors (TFRs),\cite{pike}$^-$\cite{dziedzic1} and other
metal-inorganic and -organic insulator
composites.\cite{fournier,chiteme}$^-$\cite{sun} It is clear
that, despite that many of the $t$-values reported in
Fig.\ref{fig1} are close to $t_0\simeq 2.0$, almost $50\%$ of the
measured critical exponents deviate from universality by
displaying $t\neq t_0$.

Examining all the data reported in Fig.\ref{fig1}, one observes
that the lack of universality is not limited to a particular
class of materials, although the granular metals (empty diamonds
in Fig.\ref{fig1}) have somewhat less spread values of $t$
compared to the carbon-black and TFRs composites. Another
important observation is that for the vast majority of the cases,
the nonuniversal critical exponent is {\it larger} than $t_0$,
and only few data display $t<t_0$. Finally, there is no clear
correlation between $t$ and the critical concentration value
$x_c$.

With the accumulation of experimental reports of nonuniversality,
various theories have been proposed in order to find an origin to
this phenomenon.\cite{heaney,kogut,halpe,balb,balb2} In
Ref.\onlinecite{heaney} it was argued that, in
carbon-black--polymer composites, long-range interactions could
drive the system towards the mean-field regime for which $t=t_{\rm
mf}=3.0$. This interpretation cannot however explain the
observation of critical exponents much larger than $t_{\rm mf}$,
such as those of carbon-based composites or TFRs which display
values of $t$ as high as $t\simeq 5-10$. The authors of
Ref.\onlinecite{halpe} introduced the random-void (RV) model of
continuum percolation where current flows through a conducting
medium embedding insulating spheres placed at random. By using an
earlier result,\cite{kogut} they were able to show that for this
model DC transport is described by an universality class
different from that of standard percolation model on a lattice.
The resulting critical exponent was found to be $t\simeq 2.4$ for
three dimensional systems. More recently, Balberg generalized the
RV model in an attempt to explain higher $t$-values.\cite{balb2}
The same author proposed also a model of transport nonuniversality
based on an inverted RV model in which current flows through
tunneling processes between conducting spheres immersed in an
insulating medium.\cite{balb} Within this picture, if the
distribution function of the tunneling distances decays much
slower than the tunneling decay, then the critical exponent
becomes dependent of the mean tunneling distance $a$ and in
principle has no upper bound.

In addition to the above models, there were also explications
pointing out that when Eq.(\ref{rho}) is used to fit experimental
data not restricted to the critical region, ``apparent critical
exponents'', usually larger than $t_0$, could be misinterpreted
as real critical exponents.\cite{carcia2,kolek} Although this
possibility cannot be excluded for some of the data reported in
Fig.\ref{fig1}, it is however quite unrealistic to identify the
whole set of reported nonuniversal exponents as merely apparent.

The mean-field interpretation,\cite{heaney} the RV model and its
generalization,\cite{halpe,balb2} and the tunneling inverted RV
model (also known as the tunneling-percolation model),\cite{balb}
have been devised to describe nonuniversality for various classes
of materials. For example, the RV model applies in principle to
composites where the linear size of the conducting particles is
much smaller than that of the insulating grains so that the
conducting phase can be approximated by a continuum. The
tunneling-percolation model has been instead conceived to apply to
those composites for which inter-grain tunneling is the main
microscopic mechanism of transport, such as in
carbon-black--polymer composites,\cite{sichel} or in oxide-based
TRFs.\cite{chiang,pike2,prude} In principle, the two models could
even coexist together if the continuum phase of a RV system is
made of non-sintered conducting particles interacting through
tunneling processes.

In this situation, the different proposed theories could account
for the variety of nonuniversal exponents shown in Fig.\ref{fig1}.
However, it is also true that, in order to identify a given
mechanism of nonuniversality for specific composites, little has
been done beyond a mere fit to Eq.(\ref{rho}). For example, In
Ref.\onlinecite{balb2} the DC critical exponents have been
examined together with the relative resistance noise exponent,
and in Ref.\onlinecite{chiteme} a study on AC and
magnetoresistive exponents for various composites has been
presented. As a matter of fact, no conclusive answers have been
reached and the different models of nonuniversality listed above
have not proven to really apply to real composites.

In this paper we present our contribution to the understanding of
the origin of transport nonuniversality by attacking the problem
from a different point of view. In contrast with the mean-field
hypothesis,\cite{heaney} the RV model,\cite{halpe} and its
extension,\cite{balb2} the tunneling-percolation model of Balberg
predicts that the critical exponent $t$ acquires an explicit
dependence upon a microscopic variable (the mean-tunneling
distance $a$) which could be altered by a suitable external
perturbation. So, if transport nonuniversality is driven by
tunneling, it would be possible to change the value of the
transport critical exponent $t$ by applying a pressure or a
strain to the composite. Conversely, when a material belongs to
some universality class (standard percolation theory, mean-field
universality class, or the RV model) its exponent is expected to
be independent of microscopic details and an applied strain would
not change $t$.

We have applied this idea to RuO$_2$-based TFRs whose transport
properties are known to be governed by inter-grain tunneling
processes.\cite{chiang,pike2,prude} In the following of this
paper we show that the behavior of the piezoresistive response,
{\it i. e.} the change of resistivity upon applied mechanical
strain,\cite{sichel} as a function of concentration $x$ of
RuO$_2$, can be interpreted as due to a tunneling distance
dependence of the DC critical exponent $t$, as originally
proposed in Ref.\onlinecite{balb}. The paper is organized as
follows. In the next section we briefly review the
tunneling-percolation theory and the RV model and its extension.
In Sec.\ref{piezo} we describe the theory of piezoresistivity for
percolating composites and in Sec.\ref{expe} we present our
experimental results. The last section is devoted to a discussion
and to the conclusions.

\section{Models of nonuniversality}
\label{models} The RV models,\cite{halpe,balb2} and the
tunneling-percolation theory,\cite{balb} have one point in
common. They all rely on the work of Kogut and Straley who first
developed a theoretical model of nonuniversality based on
random-resistor networks.\cite{kogut} In this model, to each
neighboring couple of sites on a regular lattice it is assigned
with probability $p$ a bond with conductance $g\neq 0$ and bond
with $g=0$ with probability $1-p$. The resulting bond conductance
distribution function is then:
\begin{equation}
\label{distri1} \rho(g)=ph(g)+(1-p)\delta(g),
\end{equation}
where $\delta(g)$ is the Dirac delta function and $h(g)$ is the
distribution function of the finite bond conductances. Close to
the bond percolation threshold $p_c$, the conductivity $\Sigma$
of the network behaves as:
\begin{equation}
\label{sigma0} \Sigma=\Sigma_0(p-p_c)^t,
\end{equation}
where $\Sigma_0$ is a prefactor. In this and in the subsequent
section we distinguish the conductivity $\Sigma$ of a
random-resistor network from that of real composites
[Eq.(\ref{rho})]. When $h(g)=\delta(g-g_0)$ where $g_0$ is some
nonzero value, Eq.(\ref{distri1}) reduces to the standard
bi-modal model for which, close to the percolation threshold, the
network conductivity $\Sigma$ follows Eq.(\ref{sigma0}) with
$t=t_0\simeq 2.0$ for all three dimensional lattices. Instead, if
$h(g)$ has a power law divergence for small $g$ of the form:
\begin{equation}
\label{distri2} \lim_{g\rightarrow 0} h(g)\propto g^{-\alpha},
\end{equation}
where $\alpha\leq 1$, then universality is lost for sufficiently
large values of the exponent $\alpha$.\cite{kogut} For lattices
of dimension $D$, the resulting conductivity critical exponent
is:\cite{machta,stenull,alava}
\begin{equation}
\label{nonuni0} t=\left\{
\begin{array}{ll}
t_0 & \mbox{if} \hspace{3mm} (D-2)\nu+\frac{1}{1-\alpha}< t_0 \\
(D-2)\nu+\frac{1}{1-\alpha} &  \mbox{if}\hspace{3mm}
(D-2)\nu+\frac{1}{1-\alpha}> t_0
\end{array}
\right.,
\end{equation}
where $t_0$ is the universal value and $\nu$ is the
correlation-length exponent ($\nu=4/3$ for $D=2$ and $\nu\simeq
0.88$ for $D=3$). For $D=3$ and by using $t_0\simeq 2.0$ and
$\nu\simeq 0.88$ the value of $\alpha$ beyond which universility
is lost is $\alpha_c=1-1/(t_0-\nu)\simeq 0.107$. Only recently
Eq.(\ref{nonuni0}) has been demonstrated to be valid to all orders
of a $\epsilon=6-D$ expansion in a renormalization group
analysis.\cite{stenull}

\begin{figure}[t]
\protect
\includegraphics[scale=0.45]{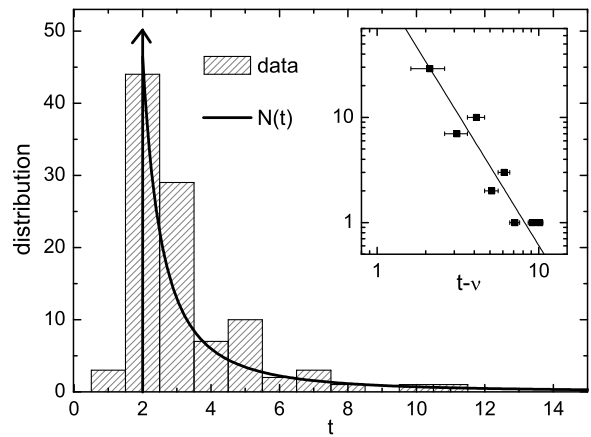}
\caption{Distribution of the $t$-values reported in
Fig.\ref{fig1}. The solid line is Eq.(\ref{distri4}) renormalized
in order to reproduce the total number of data. Inset: log-log
plot of the distribution with a fit to the power law
$a(t-\nu)^{-b}$ with $a=192\pm 62$ and $b=2.5\pm 0.4$. }
\label{fig2}
\end{figure}

In the original work of Ref.\onlinecite{kogut}, $\alpha$ was
considered as no more than a parameter of the theory without a
justification on microscopic basis. This came later with the RV
models and the tunneling-percolation
theories.\cite{halpe,balb2,balb} However, before discussing the
microscopic aspect, we find it interesting to point out that
Eq.(\ref{nonuni0}) predicts that $t$ cannot be lower than $t_0$
and it is in principle not bounded above. This is in qualitative
agreement with the experimental values of $t$ reported in
Fig.\ref{fig1}. Furthermore, Eq.(\ref{nonuni0}) can give us some
information on the distribution of $t$ values expected by the
theory. In fact, given some normalized distribution function
$f(\alpha)$ for the parameter $\alpha$ and for $D=3$, the
distribution $N(t)$ of the $t$ values is:
\begin{eqnarray}
\label{distri3} N(t)=&&\int_{-\infty}^1 \! d\alpha
f(\alpha)\delta[t-t(\alpha)]
=\delta(t-t_0)\int_{-\infty}^{\alpha_c} \! d\alpha f(\alpha)
\nonumber \\
&&+\int_{\alpha_c}^1 \! d\alpha
f(\alpha)\,\delta\!\left[t-\nu-\frac{1}{1-\alpha}\right].
\end{eqnarray}
If we assume that $f(\alpha)$ can be approximated by a constant
$f$ for $\alpha_c\leq\alpha\leq 1$, then the above expression
reduces to:
\begin{equation}
\label{distri4}
N(t)=\left(1-\frac{f}{t_0-\nu}\right)\delta(t-t_0)+
f\left(\frac{1}{t-\nu}\right)^2\theta(t-t_0),
\end{equation}
which predicts a rapid decay $N(t)\propto (t-\nu)^{-2}$ for
$t>t_0$. Equation (\ref{distri4}) is plotted in Fig.(\ref{fig2})
together with the distribution of $t$ values reported in
Fig.(\ref{fig1}). The distribution $N(t)$ has been renormalized
to the number of $t$ values and the constant $f$ has been fixed
to reproduce the number of data with $t\geq 3.0$. There is an
overall qualitative agreement between the distribution of the
experimental $t$ values and $N(t)$. In particular, the asymmetry
of the distribution and its tail for $t>t_0\simeq 2.0$ are well
reproduced. Fitting to a power law leads to a decay proportional
to $(t-\nu)^{-2.5\pm 0.4}$ (see inset of Fig.\ref{fig2}) which is
in fair agreement with the predicted behavior $(t-\nu)^{-2}$. Due
to the limited number of $t$ values available to us, the
agreement with Eq.(\ref{distri4}) could be fortuitous. However,
the point here is that the gradual decrease of the number of
times high values of $t$ are reported is not necessarily due to
``bad fits'' to Eq.(\ref{rho}),\cite{carcia2,kolek} but it can
be, at least qualitatively, explained by the form of
Eq.(\ref{nonuni0}).

Let us now discuss the microscopic origin of the exponent
$\alpha$. According to the original RV model,\cite{halpe} the
crucial parameter is the conducting channel width $\delta$ left
over from neighboring insulating spheres. For $D=3$, the
cross-section of the conducting channel has roughly the shape of
a triangle and the resulting channel conductance $g$ scales as
$g\propto \delta^{3/2}$.\cite{halpe} Hence, if $p(\delta)$ is the
distribution function of the channel width $\delta$, the
distribution $h(g)$ of the conductances reduces to:
\begin{equation}
\label{distri5} h(g)=\int d\delta
p(\delta)\delta(g-g_0\delta^{3/2}),
\end{equation}
where $g_0$ is a proportionality constant. For random
distribution of equally sized spheres, $p(\delta)$ is a constant
for $\delta \rightarrow 0$ and Eq.(\ref{distri5}) gives
$h(g)\propto g^{-1/3}$ for $g\rightarrow 0$.\cite{halpe} $h(g)$ is
therefore of the same form of Eq.(\ref{distri2}) with
$\alpha=1/3>\alpha_c$ which, according to Eq.(\ref{nonuni0}),
predicts a critical exponent $t=\nu+\frac{3}{2}\simeq 2.38$.
Despite that this value is only slightly larger than the
universal exponent $t_0\simeq 2.0$, the RV model has the merit of
being a fully microscopic justification of Eq.(\ref{distri2}). In
order to allow for higher values of $t$, in
Ref.\onlinecite{balb2} it has been proposed to relax the
condition that $p(\delta)$ is a constant for $\delta \rightarrow
0$ by introducing the more general condition $p\propto
\delta^{-\omega}$ where $\omega < 1$. By using Eq.(\ref{distri5})
one readily finds that $h(g)$ is again given by
Eq.(\ref{distri2}) but with
$\alpha=\frac{1}{3}+\frac{2}{3}\omega$.\cite{balb2} Hence, the
critical exponent is $t=t_0$ for
$\omega<\omega_c=\frac{3}{2}\alpha_c+\frac{1}{3}\simeq 0.4$ or
$t=\nu+\frac{3}{2}/(1-\omega)$ for $\omega>\omega_c$, {\it i.
e.}, $t$ is no bounded above. The use of $p\propto
\delta^{-\omega}$ has been justified to be a simple ansatz to
describe real composites in which correlations between the
conducting and insulating phases may yield to deviations from the
ideal RV system.

It is important to stress that the RV model of
Ref.\onlinecite{halpe} does not give a breakdown of universality,
but rather to a universality class different from that of the
standard DC transport percolation on a lattice. In particular,
the exponent $t=\nu+\frac{3}{2}$ does not depend on the
conductivity of the continuum and an applied isotropic strain or
pressure, albeit affecting the channel width $\delta$, does not
change the relation $g\propto \delta^{3/2}$. This should remain
true also for the generalized RV theory of Ref.\onlinecite{balb2}.

Contrary to the RV models, the tunneling-percolation theory of
nonuniversality allows for a transport critical exponent
dependent of the microscopic conductivites.\cite{balb} In this
model, current flows through tunneling processes between
neighboring conducting spheres dispersed in an insulating medium
and, for sufficiently low concentrations of the conducting
spheres, the ensemble of tunneling bonds form a percolating
network. The coexistence of tunneling and percolation has been
recently settled by experiments probing the local electrical
connectivity of various disordered systems, and a recent review on
this issue can be found in Ref.\onlinecite{balbreview}. In what
follows, we consider the situation in which grain charging
effects can be neglected with respect to the tunneling processes,
as encountered in systems with sufficiently large conducting
grains and/or high temperatures. Let us consider then the
inter-particle tunneling conductance:
\begin{equation}
\label{tunnel1} g=g_0 e^{-2(r-\Phi)/\xi},
\end{equation}
where $g_0$ is a constant, $\xi$ is the tunneling factor which is
of the order of few nm, $r$ is the distance between the centers
of two neighboring spheres of diameter $\Phi$ ($r\geq \Phi$ for
impenetrable particles). If we denote with $P(r)$ the
distribution function of adjacent inter-sphere distances $r$, the
bond conductances are then distributed according to:
\begin{equation}
\label{tunnel2} h(g)=\int_\Phi^\infty dr P(r)\,\delta\!\left[g-g_0
e^{-2(r-\Phi)/\xi}\right].
\end{equation}
Balberg notices that if $P(r)$ had a slower decay than
Eq.(\ref{tunnel1}) for $r\rightarrow \infty$, then $h(g)$ would
develop a divergence for $g\rightarrow 0$. For example, by
assuming that
\begin{equation}
\label{tunnel3} P(r)=\frac{e^{-(r-\Phi)/(a-\Phi)}}{a-\Phi},
\end{equation}
is a good approximation for the (normalized) distribution of
inter-particle distances, then Eq.(\ref{tunnel2}) would reduce to
\begin{equation}
\label{tunnel4}
h(g)=\frac{1-\alpha}{g_0}\left(\frac{g}{g_0}\right)^{-\alpha}
\end{equation}
with $\alpha=1-\frac{\xi/2}{a-\Phi}$ where $a$ is the mean
distance between neighboring particles. For
$\frac{\xi/2}{a-\Phi}\leq 1-\alpha_c\simeq 0.9$, transport
becomes nonuniversal with $t=\nu+2\frac{a-\Phi}{\xi}$.\cite{balb}
It is important to stress that here nonuniversality is not driven
by geometrical factors as in the RV model, but rather by physical
parameters such as $\xi$ and $a$. These can be different
depending on the composite and can be modified by a suitable
external perturbation. In fact, in the case of a applied pressure
or strain, the mean tunneling distance $a$ would change leading
to a modification of the critical exponent $t$. As already
pointed out in the introduction, the detection of such an effect
would be a direct signature of a tunneling-percolation--like
mechanism of transport nonuniversality.

The tunneling-percolation theory in its original formulation
relies on Eq.(\ref{tunnel3}) which should be regarded as a
phenomenological model of the distribution function of
inter-particle distances. Recently, however, a microscopic
derivation of Eq.(\ref{tunnel3}) has been formulated for a
bond-percolation regular network in which the bonds have
probability $p$ of being occupied by a string of $n$
non-overlapping spheres.\cite{grima0} In this case, in fact,
Eq.(\ref{tunnel3}) is the exact distribution function for spheres
placed in a one-dimensional channel,\cite{torquato} and the
resulting bond conductance distribution is proportional to
$g^{-\alpha_n}$ where $\alpha_n=1-\frac{\xi/2}{a_n-\Phi}$ and
$a_n$ is the mean inter-particle distance for a bond occupied by
$n$ spheres. Despite of its over-simplification, this
construction shows however that the tunneling-percolation
mechanism of nonuniversality can be justified by a fully defined
model, without need of phenomenological forms of $P(r)$.

\section{theory of piezoresistivity}
\label{piezo}

The sensitivity of the tunneling-percolation mechanism of
nonuniversality to variations of the mean tunneling distance can
be exploited by imposing a volume compression or expansion to the
system. As we discuss below, under these circumstances the
relative change of resistivity, {\it i. e.} the piezoresistive
response, changes dramatically depending on whether the DC
exponent is universal ($t=t_0$) or instead it is driven away from
universality by the tunneling-percolation mechanism.

Let us consider the rather general situation in which a
parallelepiped with dimensions $L_x$, $L_y$ and $L_z$ is
subjected to a deformation along its main axes $x$, $y$, and $z$.
The initial volume $V=L_xL_yL_z$ changes to $V(1+\phi)$ where
$\phi=\varepsilon_x+\varepsilon_y+\varepsilon_z$ is the volume
dilatation and $\varepsilon_i=\delta L_i/L_i$ are the principal
strains along $\varepsilon_i$ with $i=x,y,z$. In the absence of
strain, we assume that the conductivity $\Sigma$ of the
parallelepiped is isotropic, so that the conductance $G_i$
measured along the $i$ axis is $G_i=\Sigma L_jL_k/L_i$. For small
$\varepsilon_i\neq 0$ ($i=x,y,z$), the conductance variation
$\delta G_i$ is therefore:
\begin{equation}
\label{G1} \frac{\delta G_i}{G_i}=\frac{\delta
\Sigma_i}{\Sigma}-\varepsilon_i+\varepsilon_j+\varepsilon_k,
\end{equation}
where
\begin{equation}
\label{G2} \frac{\delta \Sigma_i}{\Sigma}=-\Gamma_\parallel
\varepsilon_i-\Gamma_\perp (\varepsilon_j+\varepsilon_k)
\end{equation}
are the relative variation of the conductivity along the
$i=x,y,z$ directions. The coefficients $\Gamma_\parallel$ and
$\Gamma_\perp$ are the longitudinal and transverse piezoresistive
factors defined as:
\begin{eqnarray}
\label{G3} \Gamma_\parallel&=& -\frac{d\ln
(\Sigma_i)}{d\varepsilon_i}=\frac{d\ln
(\rho_i)}{d\varepsilon_i} \\
\Gamma_\perp&=&-\frac{d\ln (\Sigma_j)}{d\varepsilon_i}=\frac{d\ln
(\rho_j)}{d\varepsilon_i} \,\,\,\,\,\,\,\,\, (i\neq j),
\end{eqnarray}
where $\rho_i=\Sigma_i^{-1}$ is the resistivity along the
$i$-axis and $\ln$ is the natural logarithm. The distinction
between longitudinal, $\Gamma_\parallel$, and transverse,
$\Gamma_\perp$, piezoresistive responses is important whenever
the values of the strains $\varepsilon_i$ depend on the
direction, as it is encountered when the sample is subjected to
uniaxial distortions as those induced in cantilever beam
experiments (see next section).

In what follows, we are mainly interested in the isotropic (or
hydrostatic) piezoresistive factor $\Gamma$ defined as the
resistivity change induced by the isotropic strain field
$\varepsilon_i=\varepsilon$ for all $i=x,y,z$:
\begin{equation}
\label{G4}
\Gamma=-\frac{d\ln (\Sigma)}{d\varepsilon}=\frac{d\ln
(\rho)}{d\varepsilon},
\end{equation}
which can be obtained by applying an hydrostatic pressure to the
parallelepiped. However, $\Gamma$ can also be obtained by setting
$\varepsilon_i=\varepsilon$ in Eq.(\ref{G2}) yielding:
\begin{equation}
\label{G5} \Gamma=\Gamma_\parallel+2\Gamma_\perp,
\end{equation}
which is a useful relation when the experimental set up does not
permit to apply an isotropic strain field.

Let us now study how the tunneling-percolation theory of
non-universality affect the piezoresistive factor $\Gamma$. To
this end, we assume that a cubic bond-percolation network  is
embedded in a homogeneous elastic medium and that the elastic
coefficients of the network and the medium are equal. Under an
isotropic strain field $\varepsilon_i=\varepsilon$ ($i=x,y,z$)
the mean tunneling distance $a$ changes to $a(1+\varepsilon)$
independently of the bond orientation. Hence, by assuming for
simplicity that $\Phi\rightarrow \Phi(1+\varepsilon)$, the
tunneling parameter $\alpha=1-\frac{\xi/2}{a-\Phi}$ entering
Eq.(\ref{nonuni0}) becomes $\alpha\rightarrow
\alpha+(1-\alpha)\varepsilon$ for $\varepsilon\ll 1$. According
to the discussion of Sec.\ref{models} and to
Eqs.(\ref{sigma0},\ref{nonuni0},\ref{G4}), close to the
percolation threshold $p_c$, the piezoresistive factor behaves
therefore as:
\begin{equation}
\Gamma= \left\{\begin{array}{lcc}
\Gamma_0 & \hspace{5mm} & \alpha \leq\alpha_c \\
\Gamma_0-\frac{\displaystyle dt}{\displaystyle d\varepsilon}\ln
(p-p_c) & \hspace{5mm} & \alpha >\alpha_c
\end{array}
\right., \label{G6}
\end{equation}
where
\begin{eqnarray}
\label{Gamma0} \Gamma_0&=&-\frac{d\ln(\Sigma_0)}{d\varepsilon}=-(1-\alpha)
\frac{d\ln(\Sigma_0)}{d\alpha}, \\
\label{dtde}
\frac{dt}{d\varepsilon}&=&\frac{1}{1-\alpha}=2(a-\Phi)/\xi,
\end{eqnarray}
where we have used $d\alpha/d\varepsilon=(1-\alpha)$. The
tunneling distance dependence of the DC transport exponent is
therefore reflected in a \textit{logarithmic} divergence of
$\Gamma$ as $p\rightarrow p_c$. Instead, when $\alpha\leq
\alpha_c$, the DC exponent remains equal to $t_0$ also when
$\varepsilon\neq 0$ and the resulting piezoresistive factor is
simply equal to $\Gamma=\Gamma_0$, independently of the bond
probability $p$.

It is worth to note that, as shown in Ref.\onlinecite{grima3},
contrary to $\Gamma$, the breakdown of universality has no effect
on the piezoresistive anisotropy defined as
$\chi=(\Gamma_\parallel-\Gamma_\perp)/\Gamma_\parallel$ which
behaves as $\chi\propto(p-p_c)^\lambda$ where the critical
exponent $\lambda$ is independent of $\alpha$ also when
$\alpha>\alpha_c$.\cite{grima3}

Equation (\ref{G6}) is an exact result as long as we are
concerned with the $p-p_c$ dependence close to the percolation
threshold. However, in addition to the prefactor of the logarithm,
$\Gamma$ depends also on the tunneling parameter $\alpha$ through
the term $\Gamma_0$. This dependence is far to be trivial. In
fact, consider a tensile strain ($\varepsilon>0$) which enhances
the bond tunneling resistances leading to an overall enhancement
of the sample resistivity. In this case,
$\Gamma=d\ln(\rho)/d\varepsilon$ must be strictly positive. This
means that $\Gamma_0>0$ when $\alpha\leq \alpha_c$ while, from
Eq.(\ref{G6}),  when $\alpha>\alpha_c$ $\Gamma_0$ does not need
to be positive to ensure $\Gamma>0$ because of the presence of
the logarithmic divergence. In the next subsections we provide
evidences that indeed $\Gamma_0$ changes sign in passing from
$\alpha\leq \alpha_c$ to $\alpha>\alpha_c$ by showing that both
the effective medium approximation and numerical calculations on
cubic lattices give negative values of $\Gamma_0$ for
nonuniversal DC transport. Together with the logarithmic
divergence of $\Gamma$ for $p\rightarrow p_c$, a negative value
of $\Gamma_0$ would be an additional signature of a
tunneling-percolation mechanism of nonuniversality.

\subsection{effective medium approximation}
\label{ema}
\begin{figure}
\protect
\includegraphics[scale=0.45]{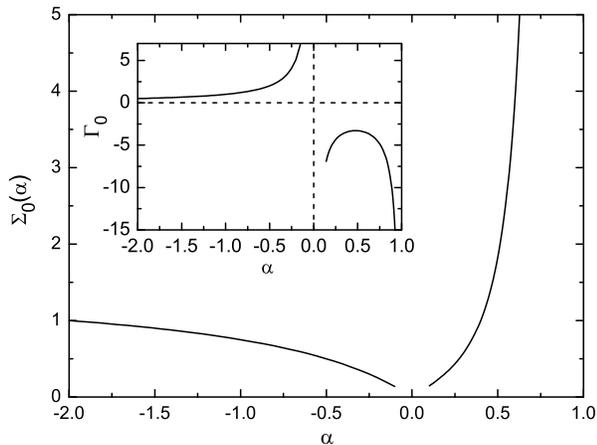}
\caption{Prefactor $\Sigma_0$ of Eq.(\ref{sigma0}) as a function
of the tunneling parameter $\alpha$. Note that $\Sigma_0$ is a
decreasing (increasing) function of $\alpha$ for $\alpha<0$
($\alpha>0$). Inset: the $p$ independent contribution $\Gamma_0$
to the piezoresistive factor $\Gamma$. Note that $\Sigma_0$ and
$\Gamma_0$ are not plotted for small values of $\alpha$ because
in this region the logarithmic corrections to $\Sigma$ calculated
within EMA change the simple power law behavior of
Eq.(\ref{sigma0}).} \label{fig3}
\end{figure}

In the effective medium approximation (EMA) the conductivity
$\Sigma=\bar{g}/\ell$ of a bond percolation cubic lattice (with
bond length $\ell$) is obtained by the solution of the following
integral equation (see for example Ref.\onlinecite{sahimi}):
\begin{equation}
\label{ema1} \int_0^\infty
dg\rho(g)\frac{\bar{g}-g}{g+2\bar{g}}=0,
\end{equation}
where $\rho(g)$ is the distribution function of the bond
conductances $g$ given in Eq.(\ref{distri1}). Close to the
percolation threshold $p_c$ ($p_c=1/3$ in EMA) and by using the
tunneling-percolation distribution of Eq.(\ref{tunnel4}) with
$g_0=1$, the above expression reduces to:
\begin{equation}
\label{ema2} (1-\alpha)\bar{g}\int_0^1 dg
\frac{g^{-\alpha}}{g+2\bar{g}}=\frac{3}{2}(p-p_c).
\end{equation}
It is clear that for $\bar{g}\rightarrow 0$ the integral in
Eq.(\ref{ema2}) remains finite as long as $\alpha<0$, while for
$\alpha>0$ it diverges as $\bar{g}^{-\alpha}$. Hence, for
$p\rightarrow p_c$  the conductivity $\Sigma$ follows the power
law behavior of Eq.(\ref{sigma0}) with $t=1$ for $\alpha<0$ and
$t=1/(1-\alpha)$ for $\alpha>0$. The corresponding prefactor
$\Sigma_0$ can be evaluated explicitly:
\begin{eqnarray}
\label{ema3a}
\Sigma_0(\alpha<0)&=&\frac{3}{2}\frac{\alpha}{1-\alpha} \\
\label{ema3b}
\Sigma_0(\alpha>0)&=&\left[\frac{3}{2^{1-\alpha}\gamma(\alpha)\gamma(2-\alpha)}\right]^{1/(1-\alpha)}
\end{eqnarray}
where $\gamma$ is the Euler gamma function. Note that the above
expressions gives $\Sigma_0(\alpha\rightarrow 0)=0$, which is a
result due to enforcing $\Sigma$ to behave as Eq.(\ref{sigma0}).
Actually, at $\alpha=0$ and for $\bar{g}\ll 1$ Eq.(\ref{ema2})
reduces to $-\bar{g}\ln(2\bar{g})=\frac{3}{2}(p-p_c)$ which leads
to logarithmic corrections in the $p-p_c$ dependence of $\bar{g}$.
For $\alpha\neq 0$, the logarithmic corrections are not important
only in a region around $p=p_c$ which shrinks to zero as
$\alpha\rightarrow 0$.

$\Sigma_0(\alpha)$ is plotted in Fig.\ref{fig3} as a function of
$\alpha$. For $\alpha<0$ ($\alpha>0$), $\Sigma_0(\alpha)$ is a
decreasing (increasing) function of $\alpha$. Hence, according to
the second equality of Eq.(\ref{Gamma0}), $\Gamma_0$ is expected
to be positive for $\alpha < 0$ and negative for $\alpha>0$. This
is confirmed in the inset of Fig.\ref{fig3} where $\Gamma_0$ is
plotted as a function of $\alpha$. Note that as
$\alpha\rightarrow 1$, $\Gamma_0$ goes to $-\infty$. In fact,
from Eq.(\ref{ema3b}), in this regime $\Sigma_0(\alpha)\simeq
\frac{1}{2}3^{1/(1-\alpha)}=\frac{1}{2}3^t$ which implies that:
\begin{equation}
\label{ema4} \Gamma_0\simeq -\ln(3)\frac{dt}{d\varepsilon}.
\end{equation}

\subsection{Monte Carlo calculations on cubic lattices}
\label{montecarlo} To evaluate the $\alpha$ dependence of the
prefactor $\Sigma_0$ of Eq.(\ref{sigma0}) we have used the
transfer-matrix method applied to a cubic lattice on $N-1$ sites
in the $z$ direction, $N$ sites along $y$, and $\cal{L}$ along the
$x$ direction.\cite{derrida} Periodic boundary conditions are used
in the $y$ direction while a unitary voltage is applied to the top
plane, and the bottom plane is grounded to zero. For sufficiently
large $\cal{L}$ ($\cal{L}\gg N$) this method permits to calculate
the conductivity $\Sigma_N$ per unit length of a cubic lattice of
linear size $N$. The transfer matrix algorithm is particularly
efficient at the bond percolation threshold $p_c\simeq
0.2488126$, and it is usually used in connection with finite size
scaling analysis of $\Sigma_N$ to extract highly accurate values
of the DC exponent $t$.\cite{normand}

In performing the calculations we have considered the following
geometries: $N=6$ (${\mathcal L}=5\times 10^7$), $N=8$
(${\mathcal L}=2\times 10^7$), $N=10$ (${\mathcal L}=1\times
10^7$), $N=12$ (${\mathcal L}=8\times 10^6$), $N=14$ (${\mathcal
L}=4\times 10^6$), and $N=16$ (${\mathcal L}=2\times 10^6$). From
Eq.(\ref{sigma0}), the resulting conductivity $\Sigma_N$ for
finite $N$ at $p=p_c$ can be written as:
\begin{equation}
\label{monte1} \Sigma_N=\Sigma_0(\alpha)[p_c-p_c(N)]^{t(\alpha)},
\end{equation}
where we have explicitly written the $\alpha$ dependence of the
prefactor and of the exponent. In the above expression, $p_c(N)$
is the percolation threshold of a finite systems of linear size
$N$. Only at $N\rightarrow \infty$, $p_c(N)$ coincides with the
percolation threshold $p_c$ of an infinite system, while for
finite values of $N$ the two quantities are related via a finite
size scaling relation of the type:\cite{gingold}
\begin{equation}
\label{monte2} p_c-p_c(N)\simeq AN^{-1/\nu}(1+BN^{-\omega}),
\end{equation}
where $\nu\simeq 0.88$ is the correlation length exponent, $A$
and $B$ are constant and $\omega$ is the first scaling correction
exponent. All the constants appearing in Eq.(\ref{monte2})
depends solely on the connectivity of the network, and are
therefore independent of the tunneling factor $\alpha$. By
inserting Eq.(\ref{monte2}) in Eq.(\ref{monte1}), the
conductivity reduces to:
\begin{equation}
\label{monte3} \Sigma_N\simeq
\overline{\Sigma}_0(\alpha)N^{-t(\alpha)/\nu}
(1+BN^{-\omega})^{t(\alpha)},
\end{equation}
where
\begin{equation}
\label{monte4}
\overline{\Sigma}_0(\alpha)=\Sigma_0(\alpha)A^{t(\alpha)}.
\end{equation}
Our strategy to calculate $\Sigma_0(\alpha)$ is the following. We
first fit or numerical data of $\Sigma_N$ with Eq.(\ref{monte3})
by setting $\omega$ fixed. In this way we obtain the exponent
$t(\alpha)$ and the prefactor $\overline{\Sigma}_0(\alpha)$. We
repeat this procedure for various values of $\alpha$ ranging from
$\alpha\simeq 1$ down to $\alpha=-\infty$ which corresponds to
the Dirac delta distribution function $h(-\infty)=\delta(g-1)$. In
this limit, $\Sigma_0(-\infty)$ is known with a rather good
accuracy, permitting us to calculate from Eq.(\ref{monte4}) the
value of the constant $A$. In this way we finally obtain
$\Sigma_0(\alpha)=\overline{\Sigma}_0(\alpha)/A^{t(\alpha)}$ by
using the values of the exponent calculated before.

\begin{figure}[t]
\protect
\includegraphics[scale=0.45]{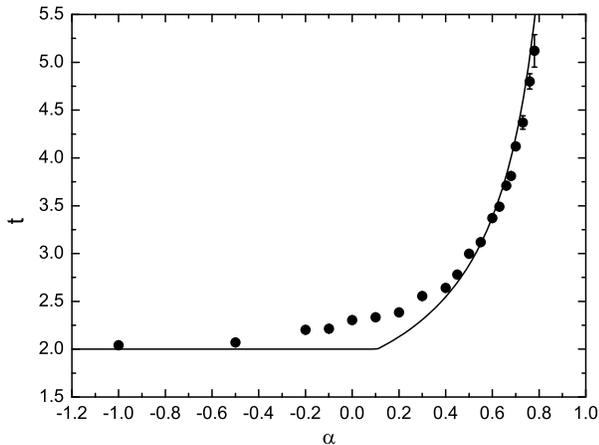}
\caption{DC transport exponent $t$ calculated from the
transfer-matrix method as a function of the parameter $\alpha$.
The solid line is the exact result Eq.(\ref{nonuni0}).}
\label{fig4}
\end{figure}
\begin{figure}
\protect
\includegraphics[scale=0.45]{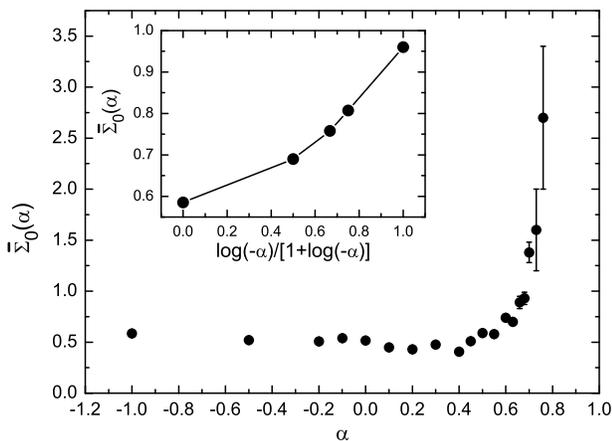}
\caption{Prefactor $\overline{\Sigma}_0(\alpha)$ of the finite
size scaling relation Eq.(\ref{monte3}) as a function of $\alpha$.
Inset: behavior of $\overline{\Sigma}_0(\alpha)$ as a function of
$\log(-\alpha)/[1+\log(-\alpha)]$ (where $\log$ is the logarithm
to base $10$) for $\alpha=-1$, $-10$, $-100$, $-1000$, and
$\alpha=-\infty$. This latter case corresponds to a Dirac delta
distribution function: $h(g)=\delta(g-1)$ for which we obtain
$\overline{\Sigma}_0(-\infty)=0.960\pm 0.007$.} \label{fig5}
\end{figure}

In Fig.\ref{fig4} we plot the values of the exponent $t$ as a
function of $\alpha$ obtained by setting $\omega=1$ in
Eq.(\ref{monte3}). We have checked that this choice for $\omega$
produced the best overall agreement with the exact result
Eq.(\ref{nonuni0}) shown in Fig.\ref{fig4}  by the solid line. In
accord with the $\alpha$-independence of $B$, we noticed little
deviations from $B\simeq 0.7$ in the whole range of $\alpha$. The
agreement between the calculated exponent and Eq.(\ref{nonuni0})
is very good far away from $\alpha=\alpha_c\simeq 0.107$. In the
vicinity of $\alpha_c$ the competition between two different
fixed points leads to a less good agreement, as already noticed
in previous works.\cite{octavio}

\begin{figure}
\protect
\includegraphics[scale=0.45]{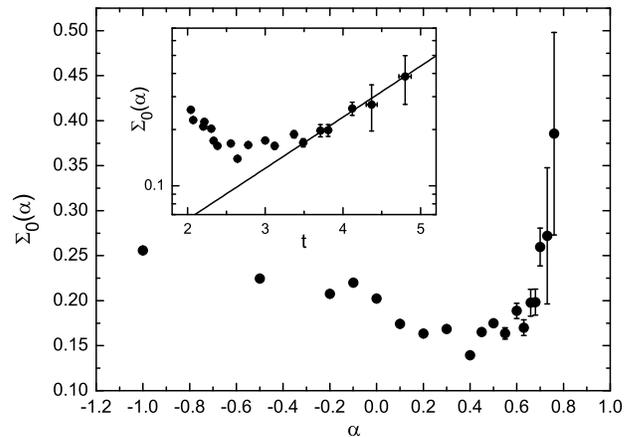}
\caption{Prefactor $\Sigma_0(\alpha)$ of the conductivity
obtained from Eq.(\ref{monte4} with $A=1.55$ and $t$ from
Fig.\ref{fig4}). Inset: semi-logarithmic plot of
$\Sigma_0(\alpha)$ as a function of the DC exponent $t$. Solid
line is a fit with $\Sigma_0(\alpha)=ab^t$ with $a=0.018\pm
0.004$ and $b=1.9\pm 0.1$.} \label{fig6}
\end{figure}

In Fig.\ref{fig5} we report the calculated values of the
prefactor $\overline{\Sigma}_0(\alpha)$ of the finite size
scaling relation Eq.(\ref{monte3}). Note that the overall
dependence of $\overline{\Sigma}_0(\alpha)$ upon $\alpha$
resembles that of Fig.\ref{fig3}, although the presence of
$A^{t(\alpha)}$ certainly affects the $\alpha> \alpha_c$ region.
In the inset of Fig.\ref{fig5} we have plotted the behavior of
$\overline{\Sigma}_0(\alpha)$ as $\alpha\rightarrow -\infty$. In
this regime, the exponent is universal ($t=t_0\simeq 2$) and the
whole $\alpha$-dependence is contained in the conductivity
prefactor $\Sigma_0(\alpha)$. From Ref.\onlinecite{clerc} we know
that $\Sigma_0(-\infty)\simeq 0.4$, while we have obtained
$\overline{\Sigma}_0(-\infty)\simeq 0.96$. Hence from
Eq.(\ref{monte4}) we have $A\simeq(0.96/0.4)^{1/2}\simeq 1.55$.

Our final results for
$\Sigma_0(\alpha)=\overline{\Sigma}_0(\alpha)/1.55^{t(\alpha)}$
are plotted in Fig.\ref{fig6} where the $t(\alpha)$ values are
those plotted in Fig.\ref{fig4}. As for the EMA case,
$\Sigma_0(\alpha)$ decreases for $\alpha$ sufficiently smaller
than $\alpha_c\simeq 0.107$ while it increases for
$\alpha>\alpha_c$. Hence, also our Monte Carlo calculations
confirm that the $p$-independent part $\Gamma_0$ of the
piezoresistive response is positive or negative, depending whether
$\alpha$ is less than or larger than $\alpha_c$, respectively. In
the inset of Fig.\ref{fig5} we report a semi-logarithmic plot of
$\Sigma_0(\alpha)$ as a function of the DC exponent $t$. For high
values of $t$, the data are reasonably well fitted by a relation
of the form $\Sigma_0(\alpha)=ab^t$ with $a=0.018\pm 0.004$ and
$b=1.9\pm 0.1$ (solid line in the inset of Fig.\ref{fig6}). Hence:
\begin{equation}
\label{monte5} \Gamma_0=-\ln(b)\frac{dt}{d\varepsilon}\simeq
-0.6\frac{dt}{d\varepsilon},
\end{equation}
confirming the asymptotic formula Eq.(\ref{ema4}) obtained within
EMA.

\section{experiment}
\label{expe}

\begin{figure}[b]
\protect
\includegraphics[scale=0.5]{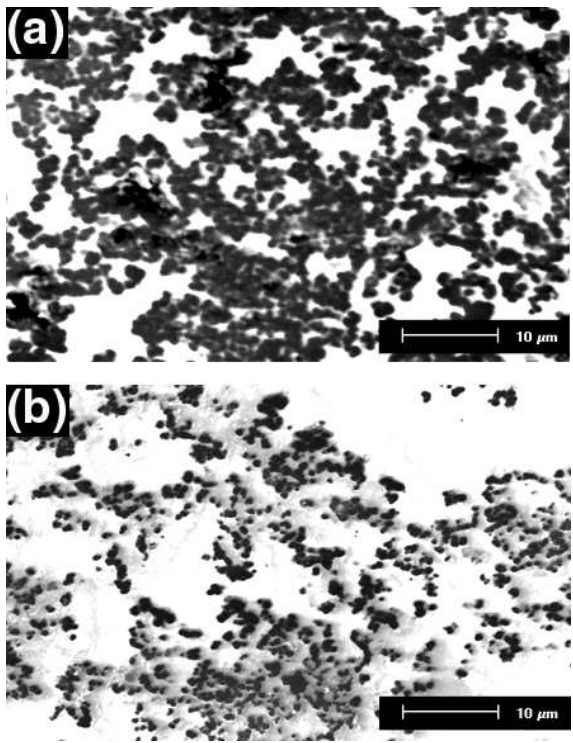}
\caption{SEM images of the surface of the A series with RuO$_2$
volume fraction $x=0.08$ and nominal RuO$_2$ grain size of $400$
nm (see text) for different firing temperatures $T_f$. (a):
$T_f=525$$^\circ$C; (b):$T_f=600$$^\circ$C} \label{fig7}
\end{figure}
\begin{figure}
\protect
\includegraphics[scale=0.5]{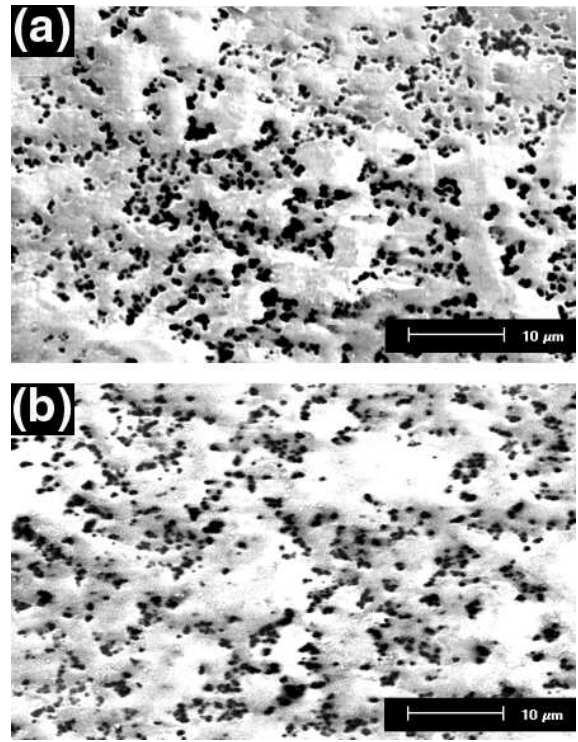}
\caption{SEM images of the surface of the B series with RuO$_2$
volume fraction $x=0.08$ and nominal RuO$_2$ grain size of $40$
nm (see text) for different firing temperatures $T_f$. (a):
$T_f=550$$^\circ$C; (b):$T_f=600$$^\circ$C} \label{fig8}
\end{figure}

In this section we describe our experiments aimed to investigate
the piezoresistive response of disordered conductor-insulator
composites made of conducting RuO$_2$ particles embedded in a
insulating glass. As shown in Fig.\ref{fig1}, this kind of TFRs
displays both universal ($t\simeq t_0$) and nonuniversal
($t>t_0$) behaviors of transport, although the factors responsible
for such changes have not yet been identified. In addition,
transport in such kind of materials is known to be governed by
electron tunneling through the glassy film separating two
neighboring conducting particles.\cite{chiang,pike2,prude} Hence,
RuO$_2$-based TFRs are ideal systems to test whether a
tunneling-percolation mechanism of transport nonuniversality sets
in.

Our samples were prepared starting with a glass frit with the
following composition: PbO ($75$\% wt), B$_2$O$_3$ ($10$\% wt),
SiO$_2$ ($15$\% wt). In order to avoid crystallisation, $2$\% wt
of Al$_2$O$_3$ was added to the glass powder. After milling, the
glass powder presented an average grain size of about $3$ $\mu$m,
as measured from laser diffraction analysis. Thermogravimetric
measurements showed negligible loss in weight, indicating glass
stability and no PbO evaporation during firing up to $800$
$^\circ$C, and differential scanning calorimetry measurements
indicated a softening temperature of about $460$ $^\circ$C. For
the conductive phase we used two different RuO$_2$ powders with
nominal grain sizes of $400$ nm (series A)and $40$ nm (series B).
Transmission electron microscope analysis confirmed that the
finer powder was made of nearly spherical particles with a
diameter of about $40$ nm, while the coarser powder had more
dispersed grain sizes ($100$ nm mean) with less regular shape.

TFRs were then prepared by mixing several weight fractions of the
two series of RuO$_2$ powder with the glass particles. An organic
vehicle made of terpineol and ethyl cellulose was added in a
quantity of about $30$\% of the total weight of the RuO$_2$-glass
mixture. The so-obtained pastes were screen printed on $96$\%
alumina substrates on pre-fired gold terminations. For the
conductance measurements, eight resistors $1.5$ mm wide and
different lengths ranging from $0.3$ mm up to $5$ mm were printed
on the same substrate. The resistors were then treated with a
thermal cycle consisting of a drying phase ($10$ min at $150$
$^\circ$C) followed by a plateau, reached at a rate of $20$
$^\circ$C/min, of $15$ min at various firing temperatures $T_f$
(see belowand Table \ref{table1}). After firing, the thickness of
the films was about $10$ $\mu$m.

In Fig.\ref{fig7} and Fig.\ref{fig8} we show scanning electron
microscope (SEM) images of the surfaces of A and B series,
respectively, with RuO$_2$ volume concentration $x=0.08$. The
firing temperature $T_f$ is $T_f=525$$^\circ$C and
$T_f=600$$^\circ$C for the A series [Figs. \ref{fig7}(a) and (b),
respectively], while $T_f=550$$^\circ$C and $T_f=600$$^\circ$C for
the B series [Figs. \ref{fig8}(a) and (b), respectively]. In the
images, dark areas indicate highly conducting regions rich of
RuO$_2$ clusters, while the white zones are instead made of
insulating glass. The grey regions surrounding the RuO$_2$
clusters indicate that some conduction is present, although much
lower than the dark areas, which we ascribe to finer RuO$_2$
particles dispersed in the glass.\cite{adachi} At the length scale
shown in the figures the conducting and insulating phases are not
dispersed homogeneously, with the RuO$_2$ clusters segregated
between large glassy regions of few micrometers in size. This
segregation effect in TFRs is well known and it is due to the
large difference in size between the fine conducting particles
and the much coarser glassy grains employed in the preparation of
the resistors.\cite{pike} Concerning the effect of the firing
temperature, it is interesting to note that for the B series
there is not much qualitative difference in the microstructure
between $T_f=550$$^\circ$C [Fig.\ref{fig8}(a)] and
$T_f=600$$^\circ$C [Fig.\ref{fig8}(b)], while for the A series it
appears that the conducting phase is more clustered at low firing
temperature [Fig.\ref{fig7}(a)] than at high $T_f$
[Fig.\ref{fig7}(b)], where the appearance of grey regions
indicate larger RuO$_2$ dispersion in the glass.

\begin{figure}[t]
\protect
\includegraphics[scale=0.42]{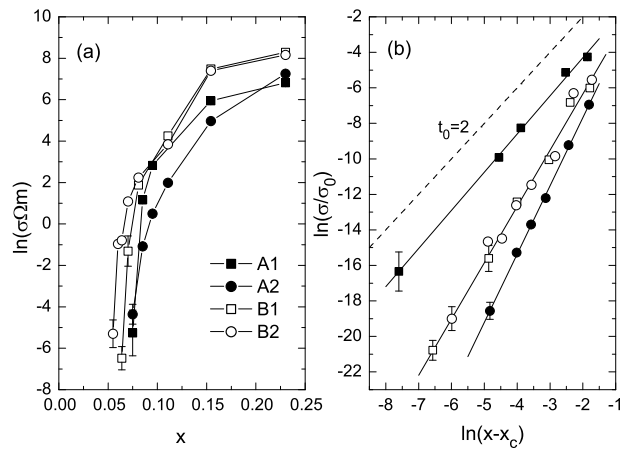}
\caption{(a): conductivity $\sigma$ as a function of RuO$_2$
volume concentration $x$ for four different series of TFRs. (b)
ln-ln plot of the same data of (a) with fits to Eq.(\ref{rho})
shown by solid lines. The dashed line has slope $t_0=2$
corresponding to universal behavior of transport. The prefactor
$\rho_0$, critical concentration $x_c$ and transport exponent $t$
values obtained by the fits are reported in Table \ref{table1}}
\label{fig9}
\end{figure}

In Fig.\ref{fig9} we report the room temperature conductivity
$\sigma$ measured for four different series of TFRs (see Table
\ref{table1}) as functions of the RuO$_2$ volume concentration
$x$. As shown in Fig.\ref{fig9}(a), $\sigma$ vanishes at rather
small values of $x$, as expected when the mean grain size of the
conducting phase ($40$ nm and $<400$ nm) is much smaller than that
of the glass (1-5 $\mu$m).\cite{kusy} The same data are
re-plotted in the ln-ln plot of Fig.\ref{fig9}(b) together with
the corresponding fits to Eq.(\ref{rho}) (solid lines) and the
best-fit parameters $\sigma_0$, $x_c$ and $t$ are reported in
Table \ref{table1} . As it is clearly shown, our conductivity data
follow the power law behavior of Eq.(\ref{rho}) with exponent $t$
close to the universal value $t_0\simeq 2.0$ for the A1 series
($t=2.15\pm 0.06$) or markedly nonuniversal as for the A2 case
which displays $t=3.84\pm 0.06$. The B1 and B2 series have nearly
equal values of $t$ ($t\simeq 3.16$) falling in between those of
the A1 and A2 series.

It is tempting to interpret the different transport behaviors of
the A1 and A2 series by referring to the microstructures reported
in Fig.\ref{fig7}. It appears that universal behavior is found
for the more clustered samples [Fig.\ref{fig7}(a)] while the
nonuniversal behavior is observed when the conducting phase is
more dispersed in the glass [Fig.\ref{fig7}(b)]. This
interpretation is coherent with the nonuniversality of both B1
and B2 series, which indeed display a large amount of RuO$_2$
dispersion evidenced by the grey regions in Figs.\ref{fig8}(a)
and (b). As discussed in Sec.\ref{models}, the microstructure has
a primary role for the onset of nonuniversality. This is
certainly true for the tunneling-percolation model in which the
microstructure governs the tunneling distribution function. In
this respect, the SEM images reported in Figs.\ref{fig7} and
\ref{fig8} may suggest that for the A1 series [Fig.\ref{fig7}(a)],
since the RuO$_2$ grains are less dispersed, the tunneling
distribution function is much narrower than those of the other
series.

\begin{table*}
\caption{\label{table1}Label legend of the various sample used in
this work with fitting parameters of Eqs.(\ref{rho},\ref{fit}) }
\begin{ruledtabular}
\begin{tabular}{cccccccccc}
Label & RuO$_2$ grain size & firing temperature $T_f$ & $x_c$ &
$\ln(\sigma_0\Omega{\rm m})$ & $t$
 & $\Gamma_0$ & $dt/d\varepsilon$  \\
\hline A1 & 400nm & $525^\circ$C  & $0.0745$ & $11.1\pm 0.3$ &
$2.15\pm 0.06$ & $16.5\pm 4.5$ & $-0.6\pm 1.2$  \\
A2 & 400nm & $600^\circ$C  & $0.0670$ & $14.2\pm 0.2$ & $3.84\pm
0.06$ & $-26.4\pm 4.8$ & $16.2\pm 1.5$  \\
B1 & 40nm & $550^\circ$C & $0.0626$ & $14.3\pm 0.5$ & $3.17\pm
0.16$ & $-45.9\pm 9$ & $26.1\pm 2.7$  \\
B2 & 40nm & $600^\circ$C & $0.0525$ & $13.7\pm 0.7$ & $3.15\pm
0.17$ & $-57.9\pm 7.2$ & $33.0\pm 2.1$   \\
\end{tabular}
\end{ruledtabular}
\end{table*}

\begin{figure}[t]
\protect
\includegraphics[scale=0.42]{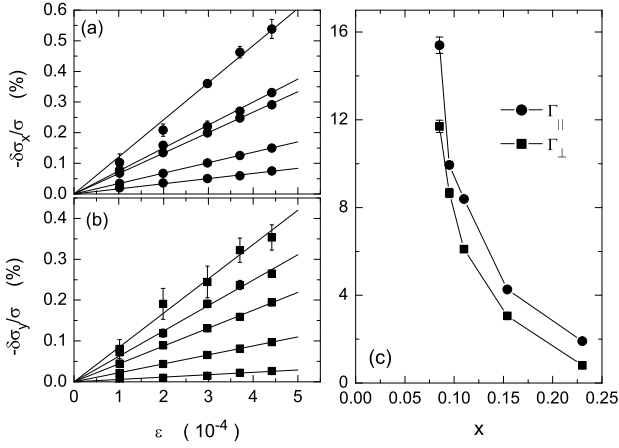}
\caption{(a) relative variation of conductivity along the $x$
axis as a function of applied strain $\varepsilon$ in cantilever
bar measurements of the A2 series for different contents $x$ of
RuO$_2$. From bottom to top: $x=0.23$, $0.154$, $0.11$, $0.095$,
and $0.085$. Solid lines are linear fits to the data. (b) the same
of (a) for the case in which the conductivity change is measured
along the $y$-axis. (c) longitudinal $\Gamma_\parallel$, and
transverse $\Gamma_\perp$, piezoresistive factors obtained by
applying Eqs.(\ref{canti1},\ref{canti2}) to the data of (a) and
(b), respectively.} \label{fig10}
\end{figure}

To measure the piezoresistive response, four resistors for each
series and with equal RuO$_2$ content were screen printed in a
Wheatstone bridge arrangement on the top of alumina cantilever
bars $51$ mm long, $b=5$ mm large, and $h=0.63$ mm thick. The
thermal treatment was the same as for the samples used to for the
conductivity measurements. The cantilever was clamped at one end
and different weights were applied at the opposite end. The
resulting substrate strain $\varepsilon$ along the main
cantilever axis can be deduced from the relation
$\varepsilon=6Mgd/(Eb h^2)$, where $d$ is the distance between
the resistor and the point of applied force, $E=332.6$ GPa is the
reduced Al$_2$O$_3$ Young modulus, $g$ is the gravitational
acceleration and $M$ is the value of the applied weight. By
fixing the main cantilever axis parallel to the $x$ direction,
then in plain strain approximation the strain field transfered to
the resistors is $\varepsilon_x=\varepsilon$, $\varepsilon_y=0$,
and $\varepsilon_z=-\nu/(1-\nu)\varepsilon$, where $\nu=0.22$ is
the Poisson ratio of $96$\% Al$_2$O$_3$. Two different
cantilevers were used for the measurements of the longitudinal
and transverse piezoresistive signals obtained by recording the
conductivity changes along the $x$ and $y$ directions,
respectively. Then, according to Eq.(\ref{G2}):
\begin{eqnarray}
\label{canti1} \frac{\delta\sigma_x}{\sigma}&=&
-\left(\Gamma_\parallel-\Gamma_\perp\frac{\nu}{1-\nu}\right)\varepsilon,
\\
\frac{\delta\sigma_y}{\sigma}&=&
-\Gamma_\perp\frac{1-2\nu}{1-\nu}\varepsilon.
 \label{canti2}
\end{eqnarray}
In Fig.\ref{fig10}(a) and Fig.\ref{fig10}(b) we plot the
conductivity variations along the $x$ and $y$ direction
respectively as a function of $\varepsilon$ for the A2 series.
The RuO$_2$ volume fractions are $x=0.23$, $0.154$, $0.11$,
$0.095$, and $0.085$ from bottom to top. In the whole range of
applied strains, the signal changes linearly with $\varepsilon$,
permitting to extract from the slopes of the linear fits of
$\delta\sigma_i/\sigma$ {\it vs} $\varepsilon$ the values of the
longitudinal and transverse piezoresistive factors through
Eqs.(\ref{canti1},\ref{canti2}). The so obtained
$\Gamma_\parallel$ and $\Gamma_\perp$ values of the A2 series are
plotted in Fig.\ref{fig10}(c) as a function of RuO$_2$ volume
concentration $x$. The main feature displayed in
Fig.\ref{fig10}(c) is that the longitudinal and the transverse
piezoresistive factors are not much different and both seem to
diverge as $x$ approaches the percolation threshold $x_c\simeq
0.067$. The fact that $\Gamma_\parallel\sim \Gamma_\perp$ is
consistent with the vanishing of the piezoresistive anisotropy
ratio $\chi=(\Gamma_\parallel-\Gamma_\perp)/\Gamma_\parallel$ at
the percolation threshold, as discussed in the previous section.

\begin{figure}[b]
\protect
\includegraphics[scale=0.42]{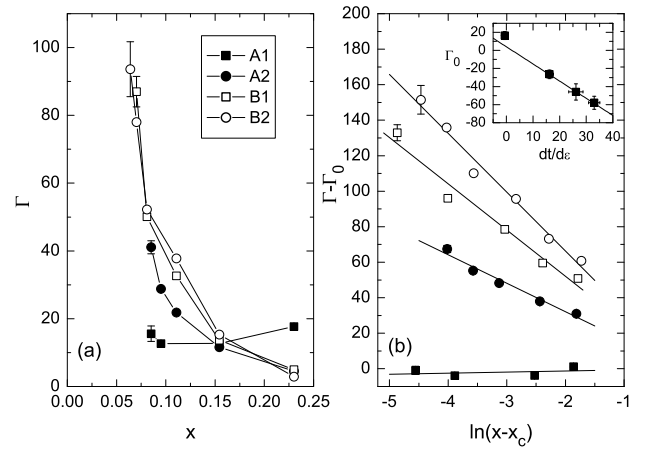}
\caption{(a) Piezoresistive factor
$\Gamma=\Gamma_\parallel+2\Gamma_\perp$ plotted as a function of
RuO$_2$ volume concentration $x$. (b) $\Gamma$ as a function of
$\ln(x-x_c)$ and fits (solid lines) to Eq.(\ref{fit}). The fit
parameters $dt/d\varepsilon$ and $\Gamma_0$ are reported in Table
I and in the inset, together with $\Gamma=-1.9dt/d\varepsilon$
(solid line).} \label{fig11}
\end{figure}

Note instead that the transport exponent for the A2 series is
$t\simeq 3.84$ (se Table \ref{table1}), that is much larger than
$t_0=2$. Hence the divergence of $\Gamma_\parallel$ and
$\Gamma_\perp$ as $x\rightarrow x_c$ could well be the signature
of a tunneling-percolation mechanism of nonuniversality. In order
to investigate this possibility we plot in Fig.\ref{fig11}(a) the
isotropic piezoresistive factor
$\Gamma=\Gamma_\parallel+2\Gamma_\perp$ for the A1, A2, B1, and
B2 series as a function of $x$. With the exception of the A1
series which displays an almost constant piezoresistive response,
the other series clearly diverge at the same critical
concentration $x_c$ at which $\sigma$ goes to zero. According to
the tunneling-percolation theory, $\Gamma$ should follow
Eq.(\ref{G6}) which predicts a logarithmic divergence at the
percolation threshold when the exponent $t$ is nonuniversal. By
using the equivalence between Eq.(\ref{G4}) and Eq.(\ref{G5}) our
data could then be used to verify this hypothesis. This is done
in Fig.\ref{fig11}(b) where $\Gamma$ is plotted as a function of
$\ln(x-x_c)$ with $x_c$ values extracted from the conductivity
data. In the entire range of concentrations, and for all the
series A1, ...,B2, $\Gamma$ is rather well fitted by the
expression:
\begin{equation}
\Gamma= \left\{\begin{array}{lcc}
\Gamma_0 & \hspace{5mm} & t=t_0 \\
\Gamma_0-\frac{\displaystyle dt}{\displaystyle d\varepsilon}\ln
(x-x_c) & \hspace{5mm} & t > t_0
\end{array}
\right., \label{fit}
\end{equation}
which is Eq.(\ref{G6}) rewritten in terms of RuO$_2$ volume
concentration $x$. The A1 series, which has DC exponent $t=2.15\pm
0.06$ close to the universal value, has no $x$ dependence of
$\Gamma$, while the other series A2, B1, and B2, characterized by
nonuniversal exponents, display a logarithmic divergence of
$\Gamma$ as $x\rightarrow x_c$. This is in agreement with the
expectations of the tunneling-percolaiton theory. Furthermore, as
shown in Table \ref{table1} and in the inset of
Fig.\ref{fig11}(b), the factor $\Gamma_0$ is positive for the A1
series and negative for A2, B1, and B2, in agreement with the
results of the last section. For the latter series, $\Gamma_0$
behaves as $\Gamma_0=-(1.9\pm 0.5)dt/d\varepsilon$ (see inset of
Fig.\ref{fig11}(b)) in accord with the asymptotic relations
obtained by EMA, Eq.(\ref{ema4}), and by our numerical
calculations, Eq.(\ref{monte5}).

\begin{figure}[t]
\protect
\includegraphics[scale=0.42]{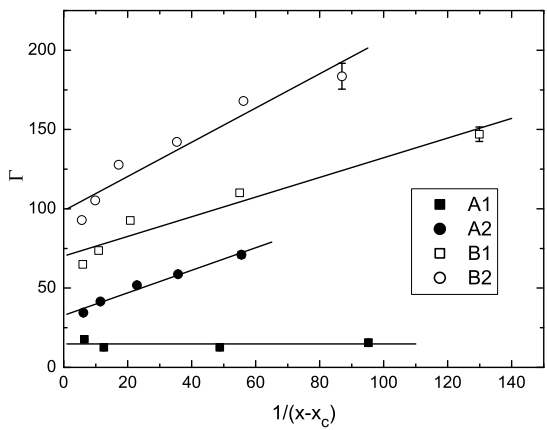}
\caption{(a) Piezoresistive factor $\Gamma$ plotted as a function
of $1/(x-x_c)$ (symbols) with tentative fits to
Eq.(\ref{gammabis}) (solid lines). The data related to different
series have been shifted vertically by $+30$ for clarity.}
\label{fig12}
\end{figure}

The logarithmic divergence of $\Gamma$ for the series having
nonuniversal values of $t$ and the corresponding negative values
of $\Gamma_0$ are features which can be coherently explained by
the tunneling-percolation model of nonuniversality. However, in
previous studies, this possibility was neglected, and the
divergence of $\Gamma$, already reported in
Ref.\onlinecite{carcia2} for TFRS and in Ref.\onlinecite{carmona2}
for carbon-black--polymer composites, was attributed to a
different mechanism independent of the universality breakdown of
$t$. This called into play the possibility of having nonzero
derivative of the volume concentration $x$ with respect to the
applied strain when the elastic properties of the conducting and
insulating phases are different. For the particular case of
RuO$_2$-based TFRs, one finds that $dx/d\varepsilon\simeq -Ax$
for small values of the RuO$_2$ concentration $x$. It is easy to
show that in this limit $A\propto 1-B_{\rm glass}/B_{{\rm
RuO}_2}$, where $B_{\rm glass}$ and $B_{{\rm RuO}_2}$ are the
bulk moduli of the glass and the conducting particles,
respectively. Since $B_{{\rm RuO}_2}\simeq 270$GPa and $B_{\rm
glass}\simeq 40-80$GPa, $A$ is expected to be different from zero
and positive. If this reasoning held true, by differentiating
Eq.(\ref{rho}) with respect to $\varepsilon$, and by keeping $t$
constant, $\Gamma=-d\ln(\sigma)/d\varepsilon$ would reduce
to:\cite{carcia2}
\begin{equation}
\label{gammabis}
\Gamma=\Gamma_0+At\frac{x}{x-x_c}=K_1+\frac{K_2}{x-x_c}
\end{equation}
where we have defined $K_1=\Gamma_0+At$ and $K_2=Atx_c$. In Fig.
\ref{fig12} we have re-plotted the $\Gamma$ values of
Fig.\ref{fig11}(a) as a function of $1/(x-x_c)$ with the same
values of the critical concentrations $x_c$ extracted from the
resistivity data. According to Eq.(\ref{gammabis}), $\Gamma$
should follow a straight line as a function of $1/(x-x_c)$ which,
although being rather correct for the A2 series, is manifestly
not true for the B1 and B2 series. In addition, the A1 series
remains almost constant, implying that $A=0$ for this case,
contrary to the premises of Ref.\onlinecite{carcia2}.

Besides from the bad fit with our data, the reasoning of
Ref.\onlinecite{carcia2} is based on a misunderstanding of the
actual physical meaning of $x$ appearing in Eq.(\ref{rho}). In
fact, $x$ should be considered just as an operative estimate of
the concentration $p$ of intergrain junctions with finite
resistances present in the sample\cite{stauffer,sahimi}. Current
can flow from one end to another of the composite as long as a
macroscopic cluster of junctions spans the entire sample. Instead
of $x$, an equally valid variable describing the integrain
junction probability $p$ would have been the concentration
\textit{in weight} of RuO$_2$, which is manifestly independent of
applied strain (and of the elastic properties of the material).

The reasoning of Refs.\onlinecite{carcia2,carmona2} is therefore
non-consistent with the physics of percolation. However, one
could tentatively argue that the applied strain actually changes
the concentration $p$ of junctions by breaking some of the bonds.
This situation could be parametrized  by allowing a $p$
dependence of $x$ so that $dx/d\varepsilon\simeq
(dx/dp)(dp/d\varepsilon)$. Also in this case however one would
end with a $1/(x-x_c)$ divergence of $\Gamma$ which we have seen
to lead to poor fits for our data (Fig.\ref{fig12}). In addition,
the values of the applied strains in our measurements are so small
($\varepsilon\sim 10^{-4}$) that their effect is that of changing
the value of the tunneling junction resistances without affecting
their concentration $p$, so that one realistically expects that
$dp/d\varepsilon=0$. This is confirmed by the results of
Fig.\ref{fig10}(a) which show no deviation from linearity for the
entire range of $\varepsilon$ values.

In a previous publication,\cite{grima1} we have re-analyzed the
piezoresitive data reported in Ref.\onlinecite{carcia2} by
assuming a logarithmic divergence of $\Gamma$ rather than the
$1/(x-x_c)$ behavior of Eq.(\ref{gammabis}). The agreement with
Eq.(\ref{fit}) was satisfactory and, furthermore, also in this
case we obtained negative values of $\Gamma_0$. However, contrary
to the present data, all the TFRs used in Ref.\onlinecite{carcia2}
were nonuniversal, so it was not possible to establish the
disappearance of the logarithmic divergence of $\Gamma$ when
$t=t_0$.

\section{conclusions}
\label{concl}

In this paper we have presented conductivity and piezoresistivity
measurements in disordered RuO$_2$-glass composites close to the
percolation threshold. We have fabricated samples displaying both
universal and nonuniversal behavior of transport with
conductivity exponents ranging from $t\simeq 2$ for the universal
samples up to $t\simeq 3.8$ for the nonuniversal ones. The
corresponding piezoresistive responses changed dramatically
depending on whether the composites were universal or not. For the
composites with $t\simeq 2$, the piezoresistive factor $\Gamma$
showed little or no dependence upon the RuO$_2$ volume fraction
$x$, whereas the nonuniversal composites displayed a logarithmic
divergence of $\Gamma$ as $x-x_c\rightarrow 0$, where $x_c$ is
the percolation threshold. We have interpreted the
piezoresistivity results as being due to a strain dependence of
the conductivity exponent when this was nonuniversal. As
discussed in Sec.\ref{piezo}, a logarithmic divergence of
$\Gamma$ is fully consistent with the tunneling-percolation model
of nonuniversality proposed by Balberg a few years ago. According
to this theory, when the tunneling distance between adjacent
conducting grains has sufficiently strong fluctuations, the DC
exponent acquires a dependence upon the mean tunneling distance
$a$. An applied strain $\varepsilon$ changes $a$ to
$a(1+\varepsilon)$ which is reflected in a
$\varepsilon$-modulation of the exponent $t$, and eventually to a
logarithmic divergence of $\Gamma=-d\ln(\sigma)/d\varepsilon$ at
$x_c$. By studying an effective medium approximation of the
tunneling-percolation model, and by extensive Monte Carlo
calculations, we have shown that when $\Gamma$ diverges, the
$x$-independent contribution $\Gamma_0$ of $\Gamma$ becomes
negative, in agreement with what we observed in the experiments.

In view of such agreement between theory and experiments, and
given the fact that an alternative explanation based on a strain
dependence of the integrain junction concentration is non-physical
and leads to poor fits to our data, we conclude that the origin of
nonuniversality in RuO$_2$-glass composites is most probably due
to a tunneling-percolation mechanism of nonuniversality. This
conclusion is also coherent with the observed correlation between
the onset of nonuniversality and the microstructure of our
samples, which showed an highly clustered arrangement of the
conducting phase when $t=t_0$ or a more dispersed configuration
when $t>t_0$. Although being only qualitative, this picture
suggests a possible route for more quantitative analysis on the
interplay between criticality and microstructure.

The tunneling-percolation mechanism of universality breakdown
could apply also to other materials for which transport is
governed by tunneling such as carbon-black--polymer composites,
and experiments on their piezoresistive response could confirm
such conjecture. Some earlier data showing diverging
piezoresistivity response at the conductor-insulator transition
already exist,\cite{carmona2,isotalo} but their interpretation is
not straightforward due to non-linear conductivity variations as
a function of strain or pressure, or even to hysteresys effects.

An interesting issue we have not addressed in the present work is
the possible effect of temperature $T$ on the piezoresistivity of
disordered composites. As shown in Ref.\onlinecite{sichel}, at
fixed concentration of the conducting phase, carbon
polyvinylchloride composites display a rather strong enhancement
of $\Gamma$ at low $T$. This was interpreted in terms of
thermally activated voltage fluctuations across the tunneling
barriers.\cite{sheng} A qualitatively similar enhancement of
$\Gamma$ as the temperature drops is expected also in models
based on variable-range hopping mechanism of
transport.\cite{grimavrh} It should be however pointed out that
in these works the problem of connectivity is not included so that
the tunneling current flows on a network without percolation
characteristics. Nevertheless, the percolation picture (with its
corresponding critical exponents) seems to survive well also in
the low temperature tunneling regime,\cite{balbreview}. Hence, a
tunneling-percolation theory of piezoresistivity at low
temperatures should be formulated by considering percolation
networks where the simple tunneling process, Eq.(\ref{tunnel1}),
are combined with additional terms describing, {\it e. g.}, grain
charging effects and/or Coulomb interactions.

This work was partially supported by TOPNANO 21 (project NAMESA,
No. 5557.2).


\begin{thebibliography}{99}
\vskip -0.5cm


\bibitem{stauffer}
D. Stauffer, and A. Aharony, {\it Introduction to Percolation
Theory} (Taylor \& Francis, London, 1994).

\bibitem{sahimi}
M. Sahimi,{\it Heterogeneous Materials I: Linear Transport and
Optical Properties} (Springer, New York, 2003).

\bibitem{batrouni}
G. G. Batrouni, A. Hansen, and B. Larson,  Phys. Rev. E {\bf 53},
2292 (1996).

\bibitem{normand}
J. -M. Normand and H. J. Herrmann, Int. J. Mod. Phys. c {\bf 6},
813 (1995).

\bibitem{clerc}
J. P. Clerc, V. A. Podolskiy, and A. K. Sarychev, Eur. Phys. J. B
{\bf 15}, 507 (2000).

\bibitem{flandin1}
L. Flandin, A. Chang, S. Nazarenko, A. Hiltner, and E. Baer, J.
Appl. Polymer Sci. {\bf 76}, 894 (2000).

\bibitem{dziedzic2}
A. Dziedzic, Inform. MIDEM {\bf 31}, 141 (2001).

\bibitem{soares}
B. G. Soares, K. M. N. Gamboa, A. J. B. Ferreira, E. Ueti, and S.
S. Camargo, J. Appl. Polymer Science {\bf 69}, 825 (1998).

\bibitem{heaney}
M. B. Heaney, Phys. Rev. B {\bf 52}, 12477 (1995).

\bibitem{flandin2}
L. Flandin, T. Prasse, R. Schueler, K. Schulte, W. Bauhofer, and
J. -Y. Cavaille, Phys. Rev. B {\bf 59}, 14349 (1999).

\bibitem{rubin}
Z. Rubin, S. A. Sunshine, M. B. Heaney, I. Bloom, and I. Balberg,
Phys. Rev. B {\bf 59}, 12196 (1999).

\bibitem{nakamura}
S. Nakamura, K. Saito, G. Sawa, and K. Kitagawa, Jpn. J. Appl.
Phys. {\bf 36}, 5163 (1997).

\bibitem{mandal}
P. Mandal, A. Neumann, A. G. M. Jansen, P. Wyder, and R. Deltour,
Phys. Rev. B {\bf 55}, 452 (1997).

\bibitem{celzard}
A. Celzard, E. Mc Rae, J. F. Mareche, G. Furdin, M. Dufort, and
C. Deleuze, J. Phys. Chem. Solids {\bf 57}, 715 (1996).

\bibitem{carmona1}
F. Carmona and C. Mourey, J. Mater. Sci. {\bf 27}, 1322 (1992).

\bibitem{carmona2}
F. Carmona, R. Canet, and S. Delhaes, J. Appl. Phys. {\bf 61},
2550 (1987).

\bibitem{czarc}
H. Czarczynska, A. Dziedzic, B. W. Licznerski, M. Lukaszewicz,
and A. Seweryn, Microelectron. J. {\bf 24} 689 (1993).

\bibitem{kubat}
J. Kubat, R. Kuzel, I. Krivka, S. Bengston, J. Prokes, and O.
Stefan, Synthetic Metals {\bf 54}, 187 (1993).

\bibitem{adriaanse}
L. J. Adriaanse, H. B. Brom, M. A. J. Michels, and J. C. M.
Brokken-Zijp, Phys. Rev. B {\bf 55}, 9383 (1997).

\bibitem{chen}
C. C. Chen and Y. C. Chou, Phys. Rev. Lett. {\bf 54}, 2529 (1985).

\bibitem{quivy}
A. Quivy, R. Deltour, A. G. M. Jansen, and P. Wyder, Phys. Rev. B
{\bf 39}, 1026 (1989).

\bibitem{putten}
D. van der Putten, J. T. Moonen, H. B. Brom, J. C. M.
Brokken-Zijp, and M. A. J. Michels, Phys. Rev. Lett. {\bf 69},
494 (1992).

\bibitem{chakra}
R. K. Chakrabarty, K. K. Bardhan, and A. Basu, Phys. Rev. B {\bf
44}, 6773 (1991).

\bibitem{hinder}
M. Hindermann-Bishoff, F. Ehrburger-Dolle, Carbon {\bf 39}, 375
(2001).

\bibitem{foulger}
S. H. Foulger, J. Appl. Polymer Sci. {\bf 72}, 1573 (1999).

\bibitem{fournier}
J. Fournier, G. Boiteux, G. Seytre, and G. Marichy, Synthetic
Metals {\bf 84}, 839 (1997).

\bibitem{pike}
G. E. Pike, in {\it Electrical Transport and Optical Properties
of Inhomogeneous Media}, edited by J. C. Garland and D. B. Tanner
(American Institute of Physics, New York, 1978) p.366.

\bibitem{angus}
H. C. Angus and P. E. Gainsbury, Electronic components {\bf 9},
84 (1968).

\bibitem{vest}
R. W. Vest, {\it conduction Mechanisms in Thick-Film
Microcircuits} Final Technical Report, ARPA Order 1642 (Purdue
Research Foundation, Ind., 1975).

\bibitem{dejeu}
W. H. de Jeu, R. W. J. Geuskens, and G. E. Pike, J. Appl. Phys.
{\bf 52}, 4128 (1981).

\bibitem{kusy0}
K. Bobran and A. Kusy, J. Phys. Condens. Matter {\bf 3}, 7015
(1991).

\bibitem{kusy1}
A. Szpytma and A. Kusy, Thin Solid Films {\bf 121}, 263 (1984).

\bibitem{kusy2}
A. Kusy, Physica B {\bf 240}, 226 (1997).

\bibitem{kusy3}
E. Listkiewicz and A. Kusy, Thin Solid Films {\bf 130}, 1 (1985).

\bibitem{kubovy}
A. Kubov\'y, J. Phys. D: Appl. Phys. {\bf 19}, 2171 (1986).

\bibitem{carcia1}
P. F. Carcia, A. Ferretti, and A. Suna, J. Appl. Phys. {\bf 53},
5282 (1982).

\bibitem{carcia2}
P. F. Carcia, A. Suna, and W. D. Childers, J. Appl. Phys. {\bf
54}, 6002 (1983).

\bibitem{dziedzic1}
A. Dziedzic, Materials Science {\bf 13}, 199 (1987).

\bibitem{chiteme}
C. Chiteme and D. S. McLachlan, Phys. Rev. B {\bf 67}, 024206
(2003).

\bibitem{lee}
S. -I. Lee, Y. Song, T. W. Noh, X. -D. Chen, and J. R. Gaines,
Phys. Rev. B {\bf 34}, 6719 (1986).

\bibitem{youngs}
I. J. Youngs, J. Phys. D: Appl. Phys. {\bf 35}, 3127 (2002).

\bibitem{abeles}
B. Abeles, H. L. Pinch, and J. I. Gittleman, Phys. Rev. Lett.
{\bf 35}, 247 (1975).

\bibitem{deptuck}
D. Deptuck, J. P. Harrison, and P. Zawadzki, Phys. Rev. Lett.
{\bf 54}, 913 (1985).

\bibitem{mamunya}
Y. P. Mamunya, V. V. Davydenko, P. Pissis, and E. V. Lebedev,
Eur. Polymer J. {\bf 38}, 1887 (2002).

\bibitem{song}
Y. Song, T. W. Noh, S. -I. Lee, and J. R. Gaines, Phys. Rev. B
{\bf 33}, 904 (1986).

\bibitem{chen2}
Y. -J. Chen, X. -Y. Zhang, T. -Y. Cai, and Z. -Y. Li, Chin. Phys.
Lett. {\bf 20}, 721 (2003).

\bibitem{maaroufi}
A. Maaroufi, K. Haboubi, A. El Amarti, and F. Carmona, J. Mater.
Sci. {\bf 39}, 265 (2004).

\bibitem{sun}
J. Sun, W. W. Gerberich, L. F. Francis, J. Polymer Sci. B {\bf
41}, 1744 (2003).

\bibitem{kogut}
P. M. Kogut and J. P. Straley,  J. Phys. C: Solid State Phys.
{\bf 12}, 2151 (1979).

\bibitem{halpe}
B. I. Halperin, S. Feng, and P. N. Sen, Phys. Rev. Lett. {\bf 54},
2391 (1985); S. Feng, B. I. Halperin, and P. N. Sen, Phys. Rev. B
{\bf 35}, 197 (1987).

\bibitem{balb}
I. Balberg,  Phys. Rev. Lett. {\bf 59}, 1305 (1987).

\bibitem{balb2}
I. Balberg, Phys. Rev. B {\bf 57}, 13351 (1998).

\bibitem{kolek}
A. Kusy and A. Kolek, Physica A {\bf 157}, 130 (1989).

\bibitem{sichel}
E. K. Sichel, P. Sheng, J. I. Gittleman, and S. Bozowski, Phys.
Rev. B {\bf 24}, 6131 (1981).

\bibitem{chiang}
Y.-M. Chiang, L. A. Silverman, R. H. French, and R. M. Cannon, J.
Am. Ceram. Soc. {\bf 77}, 1143 (1994).

\bibitem{pike2}
G. E. Pike and C. H. Seager, J. Appl. Phys. {\bf 48}, 5152 (1977).

\bibitem{prude}
C. Canali, D. Malavasi, B. Morten, M. Prudenziati, and A. Taroni,
 J. Appl. Phys. {\bf 51}, 3282 (1980).

\bibitem{machta}
J. Machta, R. A. Guyer, and S. M. Moore, Phys. Rev. B {\bf 33},
4818 (1986).

\bibitem{stenull}
O. Stenull and H.-K. Janssen,  Phys. Rev. E {\bf 64}, 056105
(2001).

\bibitem{alava}
M. Alava and C. F. Moukarzel, Phys. Rev. E {\bf 67}, 056106
(2003).

\bibitem{balbreview}
I. Balberg, D. Azulay, D. Toker, and O. Millo, Int. J. Modern
Phys. B {\bf 18}, 2091 (2004).

\bibitem{grima0}
C. Grimaldi, T. Maeder, P. Ryser, and S. Str\"assler, Appl. Phys.
Lett. {\bf 83}, 189 (2003); {\it ibid.}, Phys. Rev. B {\bf 68},
024207 (2003).

\bibitem{torquato}
S. Torquato, B. Lu, and J. Rubinstein, Phys. Rev. A {\bf 41},
2059 (1990).

\bibitem{grima3}
C. Grimaldi, T. Maeder, P. Ryser, and S. Str\"assler, Phys. Rev.
B {\bf 67}, 014205 (2003).

\bibitem{derrida}
B. Derrida and J. Vannimenus, J. Phys. A {\bf 15}, L557 (1982).

\bibitem{gingold}
D. B. Gingold and C. J. Lobb, Phys. Rev. B {\bf 42}, 8220 (1990).

\bibitem{octavio}
M. Octavio and C. J. Lobb, Phys. Rev. B {\bf 43}, 8233 (1991).

\bibitem{adachi}
K. Adachi, S. Iida, and K. Hayashi, J. Mater. Res. {\bf 9}, 1866
(1994).

\bibitem{kusy}
R. P. Kusy,  J. Appl. Phys. {\bf 48}, 5301 (1978).

\bibitem{grima1}
C. Grimaldi, T. Maeder, P. Ryser, and S. Str\"assler, J. Phys. D:
Appl. Phys. {\bf 36}, 1341 (2003).

\bibitem{isotalo}
H. Isotalo, J. Paloheimo, Y. F. Miura, R. Azumi, M. Matsumoto,
and T. Nakamura, Phys, Rev. B {\bf 51}, 1809 (1995).

\bibitem{sheng}
P. Sheng, E. K. Sichel, and J. I. Gittleman, Phys. Rev. Lett.
{\bf 40}, 1197 (1978).

\bibitem{grimavrh}
C. Grimaldi, P. Ryser, and S. Str\"assler, J. Appl. Phys. {\bf
88}, 4164 (2000).


\end{thebibliography}
\end{document}